\documentclass[letterpaper,twocolumn,english,showpacs,preprintnumbers,amsmath,amssymb]{revtex4-1}
\usepackage{float}
\usepackage{amsmath}
\usepackage{amssymb}
\usepackage{graphicx}
\usepackage{esint}

\makeatletter

\pdfpageheight\paperheight
\pdfpagewidth\paperwidth

\@ifundefined{textcolor}{}
{%
 \definecolor{BLACK}{gray}{0}
 \definecolor{WHITE}{gray}{1}
 \definecolor{RED}{rgb}{1,0,0}
 \definecolor{GREEN}{rgb}{0,1,0}
 \definecolor{BLUE}{rgb}{0,0,1}
 \definecolor{CYAN}{cmyk}{1,0,0,0}
 \definecolor{MAGENTA}{cmyk}{0,1,0,0}
 \definecolor{YELLOW}{cmyk}{0,0,1,0}
}

\usepackage{epsfig}%
\usepackage{subfigure}%
\usepackage{dcolumn}%
\usepackage{stmaryrd}%
\usepackage{mathrsfs}%
\usepackage{pifont}%
\usepackage{amsthm}%
\usepackage{bm}%
\usepackage{latexsym}%
\usepackage{amsfonts}%
\newcommand{\beq}{\begin{equation}}
\newcommand{\eeq}{\end{equation}}
\newcommand{\beqa}{\begin{eqnarray}}
\newcommand{\eeqa}{\end{eqnarray}}

\newcommand{\non}{\nonumber}

\makeatother

\begin{document}

\title{Non-Markovian quantum interference in multilevel quantum systems: Exact master equation approach}

\author{Yusui Chen$^{1,2}$}
\email{yusui.chen@stevens.edu}

\author{J. Q. You$^1$}
\email{jqyou@csrc.ac.cn}

\author{Ting Yu$^{1,2}$}
\email{ting.yu@stevens.edu}

\affiliation{ $^1$Laboratory for Quantum Optics and Quantum Information, Beijing
Computational Science Research Center, Beijing 100084, China \\
$^2$Center for Quantum Science and Engineering and Department of Physics,
Stevens Institute of Technology, Hoboken, New Jersey 07030, USA}

\date{\today}

\begin{abstract}

We study the non-Markovian quantum interference phenomenon of a multi-state atomic system coupled to a bosonic dissipative environment
by using the exact master equations derived in this paper. Two examples involving four-level systems with or without external control fields
are investigated. Our results show that non-Markovian master equations are capable of exhibiting quantum interference phenomena in a temporal domain
that has not fully explored before.  In particular, we show that the environmental memory is instrumental in the onset of non-Markovian quantum interference
pattern in different time scales.

\end{abstract}

\pacs{03.65.Yz, 03.67.Bg, 03.65.Ud, 32.90.+a}

\maketitle

\section{Introduction}

Open quantum system approach has been used extensively recently to study important quantum phenomena involving the system-environment interaction, such as
quantum decoherence, atom emission and absorption,  quantum dissipation,  quantum transport, quantum interference and quantum control,
etc. \cite{QOS1,QOS2,qc1,qc2,qc3,qc4,qc5,qi1,qi2,qd1,qd2,qd3,qdiss1}.  In the case of quantum interference, many important observations have been made for the steady states of quantum multilevel systems embedded
in a continuum medium based on a Markov master equation \cite{qc2}. With the recent advance in non-Markovian open quantum system theory,  we will be able to study the temporal behaviour of quantum
 coherence dynamics for different time scales interpolating the short-time limit and the long-time limit. However,  studies on the dynamics of quantum multilevel
 systems are technically difficult because the non-Markovian master equation for the system of interest lacks when the quantum system is coupled to an environment
 that cannot be approximated by a Markov reservoir.

The non-Markovian evolution depends on the history of the system, known as the memory effect  \cite{NM1,NM2,NM3,NM4,NM5,NM6,NM7,Ma-Yu2014}.  The non-Markovian master equations of multilevel open systems contain all the information about quantum inference patterns due to couplings to
a structured medium. Typically  the open system and its environment will
become entangled due to their interactions and the system's state will be described by a density matrix \cite{QOS1}. If the information of environmental memory
recorded in the density matrix is accessible, then the interference patterns of the quantum system will be time-dependent, and will be modified by the memory
before the system settles down to its steady state.





In recent years, the non-Markovian quantum-state diffusion equation has provided a powerful tool to study the non-Markovian dynamics of open quantum systems coupled to a bosonic environment \cite{QSD1,QSD2,QSD3,QSD4,QSD5}. However, general non-Markovian master equations are still of fundamental importance for decoherence analysis and quantum dynamics simulations \cite{NMEQ1,NMEQ2,NMEQ3,NMEQ4}. Recently, we have successfully derived a set of exact non-Markovian master equations from the non-Markovian quantum-state diffusion equation \cite{Chen-Yu2014}.
Here, we show that the temporal quantum interference for multilevel systems can also be studied rigorously in the framework of the non-Markovian master equations.

In this paper, our primary purpose is to study the quantum inference phenomena in multilevel open systems arising from many important physical processes ranging from
quantum information processing, quantum control, quantum coherence transfer, and quantum decoherence \cite{qc1,qc2,qc3,qc4,qc5,qi1,qi2,qd1,qd2,qd3,qdiss1}.  First, we systematically establish a theoretical approach to deriving the exact non-Markovian master equations to be used for
 our investigation on the temporal quantum inference.  The idea is that the time-local non-Markovian master equation for a multilevel system may be obtained
 from the corresponding non-Markovian quantum-state diffusion equation. For this purpose, we use a simple four-level atom as an example to illustrate our theoretical methods. Our major results are presented in the second example where we study the quantum interference phenomena
for a four-level atomic system controlled by the external optical pulses. In each case, we show that the exact non-Markovian master equation can be derived, and can be used to
investigate the temporal behaviors of quantum inference.

 The paper is organized as follows. In Sec. II, we will briefly review the principle ideas of quantum-state diffusion equations (QSD) approach, and show how to derive the general master equations for the multilevel systems. In Sec. III, we apply the non-Markovian master equations to investigate two examples with some brief discussions on non-Markovian features. We conclude in Sec. IV.

\section{Exact Non-Markovian Master Equations\label{ch2}}

For a generic multilevel quantum system coupled to a zero-temperature
bosonic environment, the total Hamiltonian is given by \cite{qd2} (we set $\hbar=1$)
\begin{align}
H_{{\rm tot}} & =H_{{\rm s}}+H_{{\rm int}}+H_{{\rm b}},\label{eq:Htotal}\\
H_{{\rm int}} & =L\sum_{k}g_{k}b_{k}^{\dagger}+L^{\dagger}\sum_{k}g_{k}^{*}b_{k},\nonumber \\
H_{{\rm b}} & =\sum_{k}\omega_{k}b_{k}^{\dagger}b_{k},\nonumber
\end{align}
where $L$ is the Lindblad operator of the system, describing the coupling between the system and its environment. For  a dissipative
coupling, the Lindblad operator $L$ contained in the usual form of interaction term is generally a ladder
operator in the form of $L=\sum_{n}|n\rangle\langle n+1|$, which typically satisfies the following condition:
\begin{equation}
L^{N}=0,\label{eq:close}
\end{equation}
where the integer $N$ is the number of energy levels. Note that $b_{k}^{\dagger}(b_{k})$
is the creation (annihilation) operator for the $k$th mode in the bosonic
environment and ${g_{k}}$ is the complex coupling constant between the system
and the $k$th environment mode, described by the correlation function
of the bath at zero temperature,
\begin{equation}
\alpha (t,s) = \sum_k |g_k|^2 e^{-i\omega_k (t-s)}.
\end{equation}

In the QSD approach, the dynamics of quantum systems is described by
a stochastic Schr\"{o}dinger equation. The total state $\Psi_{{\rm tot}}$
for system and environment can be expanded in the basis of Bargmann
coherent states $\Psi_{{\rm tot}}(t)=\frac{1}{\pi}\int{\rm d}^{2}z||z\rangle|\psi_{t}(z^{*})\rangle$,
where $\psi_{t}$ is the stochastic wave function for the multilevel
system. The formal QSD equation at zero temperature \cite{QSD1,QSD2}
is
\begin{align}
\partial_{t}\psi_{t}(z^{*}) & =\left(-iH_{{\rm s}}+Lz_{t}^{*}\right)\psi_{t}(z^{*})\nonumber \\
 & -L^{\dagger}\int_{0}^{t}{\rm d}s\alpha(t,s)\frac{\delta}{\delta z_{s}^{*}}\psi_{t}(z^{*}),\label{eq:QSDeq}
\end{align}
where $z_{t}^{*}=-i\sum_{k}g_{k}z_{k}^{*}e^{i\omega_{k}t}$ is the
complex Gaussian stochastic process with correlation function $\mathcal{M}[z_{t}^{*}z_{s}]=\alpha(t,s)$.
The symbol $\mathcal{M}[..]=\int\frac{{\rm d}^{2}z}{\pi}e^{-|z|^{2}}..$
means ensemble average operation over all possible stochastic trajectories
$z_{t}^{*}$. The QSD equation may be numerically solved by introducing the so-called $O$ operator:
$O(t,s)\psi_t(z^*)=\frac{\delta}{\delta z_{s}^{*}}\psi_{t}(z^{*})$, whose Markov limit is the Lindblad
operator $L$. In most cases, the $O$ operators contain complex components.  In order to clearly show the structure of $O$ operator in the basis of noise, we expand the $O$ operator and rewrite it in the form of
\begin{align}
O(t,s)= O_0 + \int_0^tds_1 z_{s_1}^*O_1(t,s,s_1)\nonumber \\
+\iint_0^tds_1ds_2 z_{s_1}^*z_{s_2}^*O_2(t,s,s_1,s_2)+... ,
\end{align}
where the terms $O_j (j=0,1,2,3...)$ are the operators according to the $j$th order of the noise. By consistency condition
$\frac{\partial}{\partial t}\frac{\delta}{\delta z_{s}^{*}}\psi_{t}=\frac{\delta}{\delta z_{s}^{*}}\frac{\partial}{\partial t}\psi_{t}$,
the $O$ operator can be formally determined by its evolution equation,
\begin{align}
\partial_{t}O & =\left[-iH_{{\rm s}}+Lz_{t}^{*}-L^{\dagger}\bar{O},\, O\right]-L^{\dagger}\frac{\delta\bar{O}}{\delta z_{s}^{*}},\label{eq:consistency}
\end{align}
where $\bar{O}(t,z^{*})=\int_{0}^{t}{\rm d}s\alpha(t,s)O(t,s,z^{*})$.
Based on the exact QSD equation \eqref{eq:QSDeq}, the reduced density
matrix $\rho_{t}$ is recovered from ensemble average over all trajectories.
The formal master equation can be written as
\begin{align*}
{\rm \partial}_{t}\rho_{t} & =-i\left[H_{{\rm s}},\,\rho_{t}\right]+L\mathcal{M}[z_{t}^{*}P_{t}]-L^{\dagger}\mathcal{M}[\bar{O}P_{t}]\\
 & +\mathcal{M}[z_{t}P_{t}]L^{\dagger}-\mathcal{M}[P_{t}\bar{O}^{\dagger}]L,
\end{align*}
where $\rho_{t}=\mathcal{M}[|\psi_{t}(z^{*})\rangle\langle\psi_{t}(z)]$.
We denote the stochastic reduced density matrix as $P_{t}=|\psi_{t}(z^{*})\rangle\langle\psi_{t}(z)|$.
By the Novikov theorem \cite{novikov}, the ensemble average can be calculated by $\mathcal{M}[z_{t}^{*}P_{t}]=\int_{0}^{t}{\rm d}s\mathcal{M}[z_{t}^{*}z_{s}]\mathcal{M}[\frac{\delta}{\delta z_{s}}P_{t}]$.
Since $|\psi_{t}\rangle$ is independent of noise $z_{t}$, we have $\mathcal{M}[\frac{\delta}{\delta z_{s}}P_{t}]=\mathcal{M}[P_{t}O^{\dagger}]$.
Then the above formal master equation can be further written as
\begin{equation}
{\rm \partial}_{t}\rho_{t}=-i\left[H_{{\rm s}},\,\rho_{t}\right]+\left[L,\,\mathcal{M}[P_{t}\bar{O}^{\dagger}]\right]-\left[L^{\dagger},\,\mathcal{M}[\bar{O}P_{t}]\right].\label{eq:fromalmeq}
\end{equation}
When the environment is Markov and the corresponding correlation function $\alpha(t,s)=\delta(t,s)$,
then $\mathcal{M}[z_{t}^{*}P_{t}]=\mathcal{M}[P_{t}L]=\rho_{t}L$, and
Eq. \eqref{eq:fromalmeq} is reduced to the master equation of the standard Lindblad
form \cite{QOS2}. However, $O$ operator generally
contains noise in non-Markovian regime, so $\mathcal{M}[P_{t}\bar{O}^{\dagger}]$
cannot reduce to a simply form.
By repeatedly using Novikov theorem, we can write the $n$th-order noise term as
\begin{align}
 & \mathcal{M}[z_{s_{1}}z_{s_{3}}...z_{s_{2n-1}}P_{t}]\nonumber \\
= & \int_{0}^{t}{\rm d}s_{2}\alpha_{1,2}\mathcal{M}\left[\left(z_{s_{3}}...z_{s_{2n-1}}\right)\frac{\delta P_{t}}{\delta z_{s_{2}}^{*}}\right]\nonumber \\
= & \int_{0}^{t}{\rm d}s_{2},..,{\rm d}s_{2n}\left(\prod_{j=1}^{n}\alpha_{j,j+1}\right)\mathcal{M}\left[\left(\prod_{j=1}^{n}\frac{\delta}{\delta z_{s_{2j}}^{*}}\right)P_{t}\right],\label{eq:novikov}
\end{align}
where $\alpha_{i,j}=\alpha(s_{i},s_{j})$. Next, by applying this
inference of Novikov theorem on $\mathcal{M}[P_{t}\bar{O}^{\dagger}]$,
we can have an infinite series of increasing power order of $O$ operator and $O^{\dagger}$ operator. Now let us recall the identity function of $O$ operator in Eq. \eqref{eq:consistency}, the right side contains up to $(N-2)^{th}$ order of noise, while for the left side, the commutator term $[L^{\dagger}\bar{O}_{j},\, O_{k}]$ generates up to $(j+k)^{th}$ order of noise. Naturally, the two sides of Eq.~\eqref{eq:consistency} match with each other and construct a set of closed differential equations, which means that $\bar{O}_{j}O_{k}$ must go to zero if $j+k>N-2$. For these disappeared terms, we name them ``forbidden conditions'' \cite{Chen-Yu2014}:
\begin{equation}
O_{j}O_{k}=0,\qquad(j+k>N-2).\label{eq:forbidden}
\end{equation}
Then it's easy to prove that the $\mathcal{M}[z_{s_{1}}z_{s_{3}}...z_{s_{2n-1}}P_{t}]$
must be a sum of finite terms. We will also have detailed
discussion in the next part. If we denote $R(t)=\mathcal{M}[P_{t}\bar{O}^{\dagger}]$,
the compact form of the exact master equation is written as
\begin{equation}
{\rm \partial}_{t}\rho_{t}=-i\left[H_{{\rm s}},\,\rho_{t}\right]+\left[L,\, R(t)\right]+\left[L,\, R(t)\right]^{\dagger}.\label{eq:exactmeq}
\end{equation}

\section{Temporal quantum interference based on the exact master equations\label{ch3}}

\subsection{Four-level atom model}

The first model we consider as an example is the four-level atom model described
by the total Hamiltonian in the same form as Eq.~\eqref{eq:Htotal},
with
\begin{align}
H_{{\rm s}} &= \sum_{m=1}^{4}\omega_{m}|m\rangle\langle m|, \nonumber\\
L &=\sum_{m=1}^{3}\kappa_{m}|m\rangle\langle m+1|,
\end{align}
where the operator $|m\rangle\langle m+1|$ is the ladder angular-momentum
operator. It is obvious that $L^3 = 0$. Following the previous discussion in Sec. II, $O$ operator contains up to second order of noise and is in the form of
\begin{align}
O(t,s) & =O_{0}(t,s)+\int_{0}^{t}{\rm d}s_{1}z_{s_{1}}^{*}O_{1}(t,s,s_{1})\nonumber \\
 & +\iint_{0}^{t}{\rm d}s_{1}{\rm d}s_{2}z_{s_{1}}^{*}z_{s_{2}}^{*}O_{2}(t,s,s_{1},s_{2}).
\end{align}
Substituting this solution into Eq. \eqref{eq:consistency}, we have
\begin{align}
\partial_{t}O_{0}(t,s) & =\left[-iH_{{\rm s}},\: O_{0}\right]-\left[L^{\dagger}\bar{O}_{0},\, O_{0}\right]-L^{\dagger}\bar{O}_{1}(t,s),\nonumber \\
\partial_{t}O_{1}(t,s,s_{1}) & =\left[-iH_{{\rm s}},\: O_{1}\right]-\left[L^{\dagger}\bar{O}_{0},\, O_{1}\right]-\left[L^{\dagger}\bar{O}_{1},\, O_{0}\right]\nonumber \\
 & -2L^{\dagger}\bar{O}_{2}(t,s,s_{1}),\nonumber \\
\partial_{t}O_{2}(t,s,s_{1},s_{2}) & =\left[-iH_{{\rm s}},\: O_{2}\right]-\left[L^{\dagger}\bar{O}_{0},\, O_{2}\right]-\left[L^{\dagger}\bar{O}_{1},\, O_{1}\right]\nonumber \\
 & -\left[L^{\dagger}\bar{O}_{2},\, O_{0}\right].\label{eq:4levelO}
\end{align}
Meanwhile we have the boundary conditions for each term of $O$ operator:
\begin{align}
\left[L,\,O_0 \right] &= O_1(t,s,t), \nonumber\\
 \left[L,\,O_1 \right] &=2 O_2(t,s,t,s_1). \nonumber
\end{align}
Next, the ensemble average term $R(t)$ in the formal master equation \eqref{eq:exactmeq}
is
\begin{align}
R(t) & =\rho_{t}\bar{O}_{0}^{\dagger}+\int_{0}^{t}{\rm d}s_{1}\alpha_{1,2}\mathcal{M}[z_{s_{1}}P_{t}]\bar{O}_{1}^{\dagger}\nonumber \\
 & +\int_{0}^{t}{\rm d}s_{1}\alpha_{1,2}\mathcal{M}[z_{s_{1}}z_{s_{2}}P_{t}]\bar{O}_{2}^{\dagger}. \label{eq:R}
\end{align}
By repeating applying Novikov's theorem \eqref{eq:novikov}, the two terms $\mathcal{M}[z_{s_{1}}P_{t}]$ and $\mathcal{M}[z_{s_{1}}z_{s_{2}}P_{t}]\bar{O}_{2}^{\dagger}$ can be extensively written in an infinite long series. However, with the criterion "forbidden conditions" \eqref{eq:forbidden}, we have
\beqa
& &\mathcal{M}[z_{s_{1}}P_{t}]\bar{O}_{1}^{\dagger} \non\\
& =&\int_{0}^{t}{\rm d}s_{2}\alpha_{1,2}O_{0}(t,s_{2})\rho_{t}\bar{O}_{1}^{\dagger}\non\\
 & +&\iiint_{0}^{t}{\rm d}s_{2}s_{3}s_{4}\alpha_{1,2}\alpha_{3,4}O_{1}(t,s_{2},s_{3})\rho_{t}O_{0}^{\dagger}(t,s_{4})\bar{O}_{1}^{\dagger},\non
 \eeqa
 and
 \beqa
& &\mathcal{M}[z_{s_{1}}z_{s_{2}}P_{t}]\bar{O}_{2}^{\dagger}\non \\
& =&\iint_{0}^{t}{\rm d}s_{3}s_{4}\alpha_{1,3}\alpha_{2,4}O_{0}(t,s_{3})O_{0}(t,s_{4})\rho_{t}\bar{O}_{2}^{\dagger}\non\\
 & +& \iint_{0}^{t}{\rm d}s_{3}s_{4}\alpha_{1,3}\alpha_{2,4}O_{1}(t,s_{3},s_{4})\rho_{t}\bar{O}_{2}^{\dagger}.\non
\eeqa
Substituting these terms into Eq. \eqref{eq:R}, we obtain that
\begin{align}
& R(t)=\rho_{t}\bar{O}_{0}^{\dagger}+\iint_{0}^{t}{\rm d}s_{1}{\rm d}s_{2}\alpha_{1,2}O_{0}(t,s_{2})\rho_{t}\bar{O}_{1}^{\dagger}(t,s_{1})\nonumber \\
& +\iiiint_{0}^{t}{\rm d}\bar{s}\alpha_{1,2}\alpha_{3,4}O_{1}(t,s_{2},s_{3})\rho_{t}O_{0}^{\dagger}(t,s_{4})\bar{O}_{1}^{\dagger}(t,s_{1})\nonumber \\
 & +\iiiint_{0}^{t}{\rm d}\bar{s}\alpha_{1,3}\alpha_{2,4}O_{0}(t,s_{3})O_{0}(t,s_{4})\rho_{t}\bar{O}_{2}^{\dagger}(t,s_{1},s_{2})\nonumber\\
 & +\iiiint_{0}^{t}{\rm d}\bar{s}\alpha_{1,3}\alpha_{2,4}O_{1}(t,s_{3},s_{4})\rho_{t}\bar{O}_{2}^{\dagger}(t,s_{1},s_{2}),
\end{align}
where $\rm{d}\bar{s}$ denotes the integral elements $\rm{d}s_{1}\rm{d}s_{2}\rm{d}s_{3}\rm{d}s_{4}$. The exact non-Markovian master equation for four-level system is determined as the same form as in Eq. \eqref{eq:exactmeq}.

In order to simplify the numerical simulation, we choose Ornstein-Uhlenbeck correlation
function, $\alpha(t,s)=\frac{\gamma}{2}e^{-\gamma|t-s|}$, for the
advantage of showing the transference between Markov $(\gamma\rightarrow\infty)$
and non-Markovian ($\gamma$ is small ) process. However, it needs to point out that, in our general
derivation, the correlation function is arbitrary.
\begin{figure}[h]
\includegraphics[scale=0.42]{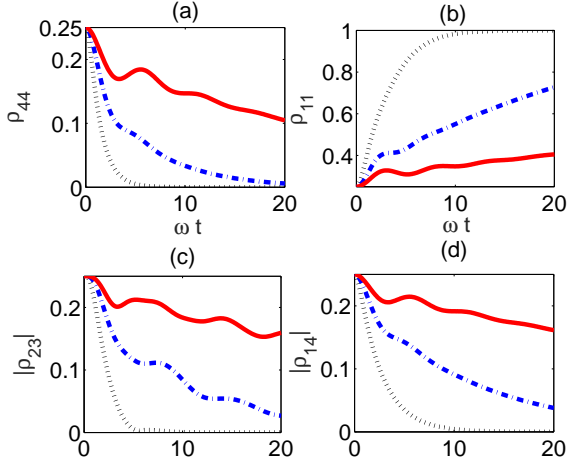}

\caption{(Color online) Non-Markovian evolution of four-level atom. The initial
state is set as $|\psi\rangle_{0}=(|1\rangle+|2\rangle+|3\rangle+|4\rangle)/2$.
We show the population of highest level (a) and ground level (b),
also the coherence of $|\rho_{23}|$ (c) and $|\rho_{14}|$ (d). The
solid (red) curve is for $\gamma=0.$2, dash-dotted (blue) is for $\gamma=0.5$
and dotted (black) is for $\gamma=2$. \label{high_spin}}
\end{figure}

As shown in Fig. \ref{high_spin}, the time evolution of population and coherence
is significantly different in Markov and non-Markovian regime. Our result
clearly reveals that the decay speed of population is much slower in non-Markovian
regime than in Markov limit, and the decay behavior of interference is similar. As expected, the revival behavior is discovered in the non-Markovian regime.

\subsection{Quantum Interference}

\begin{figure}[h]
\includegraphics[scale=0.3]{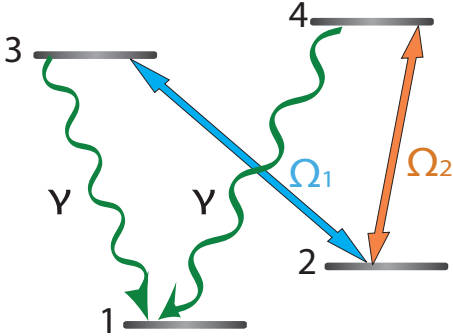}

\caption{Schematic diagram of quantum interference model, which shows the two
upper level ($|4\rangle$, $|3\rangle$) coupled to the intermediate
level ($|2\rangle$) with Rabi frequency $\Omega_1$ and $\Omega_2$, and decaying to
ground state ($|1\rangle$). \label{4-level_model}}
\end{figure}

Quantum interference in multilevel system leads to many interesting
effects, such as quantum absorption reduction and cancelation \cite{qc2}.
Here in this section, we consider a four-level system as shown in
Fig. \ref{4-level_model}. The Hamiltonian can be written as,
\begin{align*}
H_{{\rm tot}} & =H_{{\rm s}}+H_{{\rm int}}+H_{{\rm b}},\\
H_{{\rm s}} & =\sum_{m=1}^{4}\omega_{m}|m\rangle\langle m|+(\Omega_{1}e^{-i\mu t}|3\rangle\langle2|+h.c.)\\
&+(\Omega_{2}e^{-i\mu t}|4\rangle\langle2|+h.c.),\\
H_{{\rm int}} & =\sum_{k}\sum_{j=3,4}g_{k}|j\rangle\langle1|b_{k}+h.c.,\\
H_{{\rm b}} & =\sum_{k}\omega_{k}b_{k}^{\dagger}b_{k},
\end{align*}
where the $\Omega_{1(2)}$ is the Rabi frequency of the driving field for transition between
$|4(3)\rangle$ and $|2\rangle$ ( see Fig.\ref{4-level_model}). There are two decay channels for spontaneous emission. The decaying constant between
the system and $k^{th}$ mode in environment is $g_{k}$. Thus, the Lindblad
operator for this system is written as
\[
L=|1\rangle\langle3|+\kappa|1\rangle\langle4|,
\]
where $\kappa$ is the ratio of decaying constants between these two channels.
It is easy to prove that this model satisfies the ``forbidden condition''
and $L^3 =0$. Because the dissipative channels are
independent of the driving field, we can apply the results in Eq. (3) for $O$ operators and the exact master equation in \eqref{eq:exactmeq}
directly for numerical simulation to study the quantum interference under the external field control.

\begin{figure}
\includegraphics[scale=0.65]{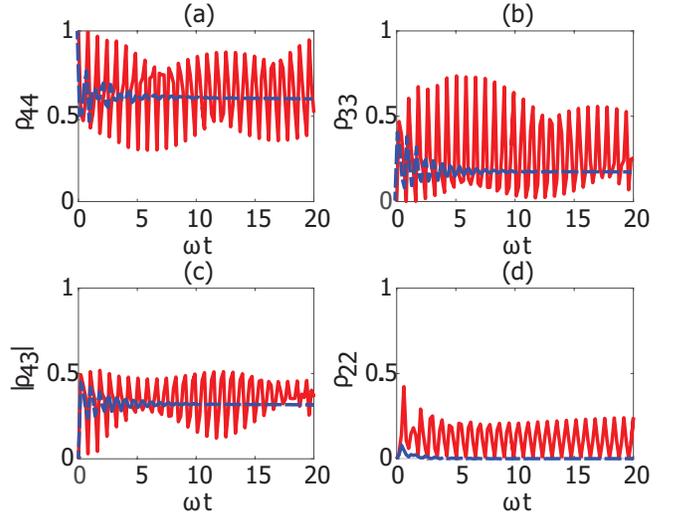}

\caption{(Color online) Time evolution of population of $|4\rangle$, $|3\rangle$,
$|3\rangle$ and coherence $|\rho_{43}|$ between $|4\rangle$ and $|3\rangle$. Initial state is set as $|4\rangle$.
The constants are set as: $\omega_4-\omega_3=\omega$, $\Omega_{1}=5\omega$,
$\Omega_{2}=2\Omega_{1}$, $\mu=2\omega,$ $\kappa=2$. Take comparison
for two $\gamma$ values. (1) $\gamma=0.5$, solid (red), (2) $\gamma=10$,
dashed (blue). \label{df_1}}
\end{figure}
\begin{figure}
\includegraphics[scale=0.65]{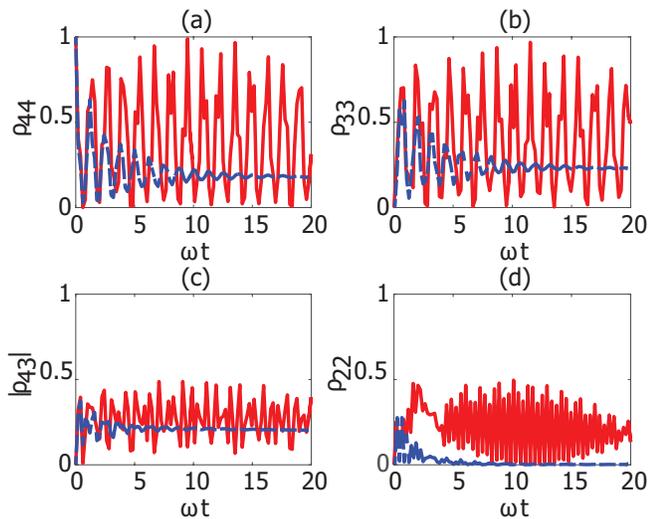}

\caption{(Color online) Time evolution of population of $|1\rangle$, $|2\rangle$,
$|3\rangle$ and coherence $|\rho_{43}|$ between $|4\rangle$ and $|3\rangle$. Initial state is set as $|4\rangle$.
The constants are set as: $\omega_4-\omega_3=\omega$, $\Omega_{1}=\Omega_{2}=5\omega$,
$\mu=2\omega$, $\kappa=1$. Take comparison for two $\gamma$ values.
(1) $\gamma=0.5$, solid (red), (2) $\gamma=10$, dashed (blue).\label{df_2}}
\end{figure}

In Figs. \ref{df_1} and \ref{df_2}, time evolution of the population
of level $2$, level $3$ and level $4$ and the coherence of the upper two levels are shown. By choosing different decay rate $\gamma$ in the Ornstein-Uhlenbeck correlation function $\frac{\gamma}{2}e^{-\gamma |t-s|}$, the short-time evolution in Markov regime and in non-Markovian regime is compared. With our exact master equation, numerical simulation shows that the population of the upper two levels are swapping in a high frequency, when $\gamma = 0.5$ (the red solid line), a typical non-Markovian regime. At the same time, the average population of the highest level $4$ fluctuates between the range from $0.4$ to $1$. While in the Markov regime, $\gamma =10$ ( blue dashed line), the population quickly decays to a steady state. Because of the real decay rate we used in the simulation, Ornstein-Uhlenbeck noise's long-time limit is same as its Markov limit. However, as we emphasized in the beginning, many devices, for instance high-Q cavities, do not satisfy the Markov environment assumption. For example, a visible light laser pulse can easily lasts in a high-Q cavity longer than 20 femtosecond scale before achieving a steady state. Otherwise, the measurement on the system would lead to a large volatility.

The second conclusion is the population inverse only happens in the level $4$ and level $3$ under some particular conditions. In Fig. \ref{df_2}, two parameters are changed, $\Gamma_2=\Gamma_1$ and $\kappa=1$. The populations of level $4$ and level $3$ of the steady state are close to each other. Although the conditions of population inverse was mentioned in \cite{qc2}, the discussion was based on the spectral density analysis and Markov assumption. Our exact master equation approach supplies a alter way to study the steady state of the system under arbitrary environmental spectrum.

\begin{figure}
\includegraphics[scale=0.65]{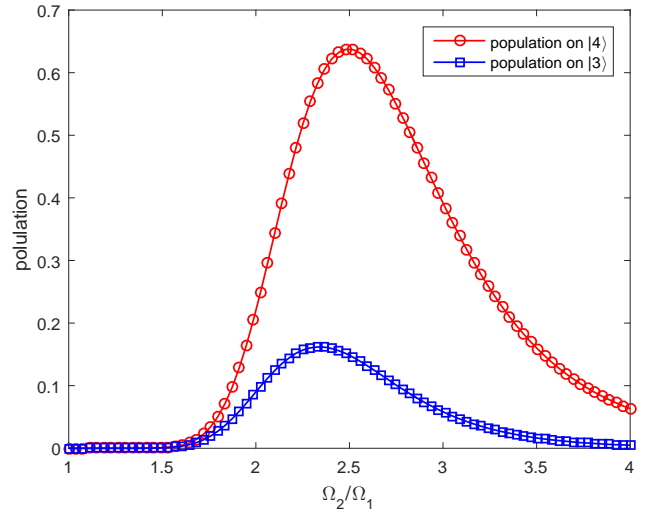}

\caption{(Color online) The population of $|4\rangle$ (red-circle), $|3\rangle$ (blue-square) at long time limit depending on the ratio of $\Omega_{2}/\Omega_{1}$. Initial state is set as $|4\rangle$. The constants are set as: $\mu=2\omega$, $\kappa=2$.\label{lll_1}}
\end{figure}

\begin{figure}
\includegraphics[scale=0.65]{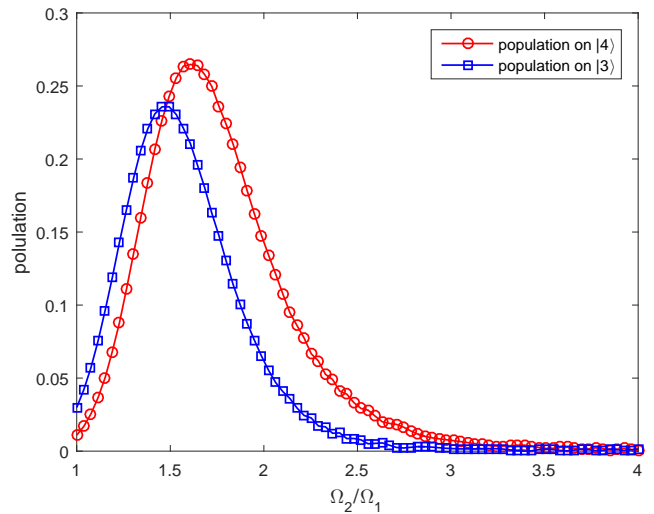}

\caption{(Color online) The population of $|4\rangle$ (red-circle), $|3\rangle$ (blue-square) at long time limit depending on the ratio of $\Omega_{2}/\Omega_{1}$. Initial state is set as $|4\rangle$. The constants are set as: $\mu=2\omega$, $\kappa=1$\label{lll_2}}
\end{figure}

In order to study the influence of the parameters to the steady state, we simulate the steady state populations of level $4$ and level $3$ under different $\kappa$ and the ratio $\Omega_2/\Omega_1$. In Fig. \ref{lll_1}, the strongest population inverse happens around the ratio $\Omega_2/\Omega_1 = 2.5$ and the population of level $4$ reaches the maximal value as 0.65. While in Fig. \ref{lll_2}, the population inverse only happens when the ratio $\Omega_2/\Omega_1 > 1.5$ and the steady populations of level $4$ and level $3$ are both close to 0.25, once $\kappa$ is changed from $2$ to $1$. With our master equation approach, the temporal control on multi-level systems can be simulated under considering any parameters in the set up and arbitrary environmental spectrum.

\section{Conclusion\label{ch4}}

In summery, we demonstrate the interesting quantum interference patterns induced by
the non-Markovian environment with the aid of the exact master equation for the multilevel systems.
Several interesting examples are used to show how quantum memory affects system's dynamics, and
in each case, we have shown that the exact master equations can play a very important role in
exploring non-Markovian open system dynamics and quantum control. One important feature
illustrated by the exact master equations in each case is that quantum interference can be
investigated in all the physically interesting scales ranging from short-time limit to the long-time
limit which is typically described by a Markov master equation. A future work along this
line of research will be the detailed study of quantum interference and quantum control of
multilevel systems coupled to  a super-ohmic or sub-ohimic environment.

\begin{acknowledgments}
We acknowledge grant support from the NSF PHY-0925174. J.Q.Y. is supported by the National Key Research and Development Program of China (Grant No. 2016YFA0301200),  the NSFC (Grant No. 11774022), and the NSAF (Grant No. U1530401).

\end{acknowledgments}


\begin{thebibliography}{10}

\bibitem{QOS1}H. P. Breuer, F. Petruccione, \emph{Theory of Open Quantum Systems} (Oxford. New York, 2002).
\bibitem{QOS2}C. W. Gardiner and P. Zoller, \emph{Quantum Noise} (Springer-Verlag, Berlin, 2004).


\bibitem{qc1} H. M. Wiseman and G. J. Milburn, Phys. Rev. Lett. \textbf{70},  548 (1993).
\bibitem{qc2} S. Y. Zhu and M. O. Scully, Phys. Rev. Lett. \textbf{76},388 (1996).
\bibitem{qc3} A. C. Doherty, S. Habib, K. Jacobs, H. Mabuchi, and S. M. Tan,  Phys. Rev. A {\bf 62}, 012105, (2000).
\bibitem{qc4} H. Rabitz, R. de Vivie-Riedle, M. Motzkus, and K. Kompa, Science \textbf{288}, 824 (2000).
\bibitem{qc5} G. S. Uhrig, Phys. Rev. Lett. \textbf{98}, 100504 (2007).

\bibitem{qi1}A. Vaziri, G. Weihs, and A. Zeilinger, Phys. Rev. Lett. {\bf 89}, 240401 (2002).
\bibitem{qi2}J. Claudon, F. Balestro, F. W. J. Hekking, and O. Buisson, Phys. Rev. Lett. {\bf 93}, 187003 (2004).

\bibitem{qd1}L. Viola, E. Knill, and S. Lloyd, Phys. Rev. Lett. \textbf{82}, 2417 (1999).
\bibitem{qd2}T. Pellizzari, S.A. Gardiner, J.I. Cirac, and P. Zoller, Phys. Rev. Lett. \textbf{75}, 3788 (1995).
\bibitem{qd3}  X. Wang, A. Miranowicz, Y. X. Liu, C. P. Sun, and F. Nori, Phys. Rev. A {\bf 81}, 022106 (2010); J. Ma, Z. Sun, X. Wang,
and F. Nori, Phys. Rev. A {\bf 85}, 062323 (2012).

\bibitem{qdiss1} M. J. Kastoryano, M. M. Wolf, and J. Eisert, Phys. Rev. Lett. \textbf{110}, 110501 (2013).




\bibitem{NM1}T. Yu and J. H. Eberly, Phys. Rev. Lett.  \textbf{93}, 140404 (2004).
\bibitem{NM2}B. L. Hu, J. P. Paz, Y. Zhang, Phys. Rev. D \textbf{45}, 2843 (1992).
\bibitem{NM3}J. P. Paz and A. J. Roncaglia, Phys. Rev. Lett. {\bf 100}, 220401 (2008).
\bibitem{NM4}W. T.  Strunz and T. Yu, Phys. Rev. A \textbf{69}, 052115 (2004).
\bibitem{NM5}T. Yu and J. H. Eberly,  Phys. Rev. B \textbf{68}, 165322 (2003).
\bibitem{NM6}J.S. Xu, C. F. Li, M. Gong, X. B. Zou, C.H. Shi, G. Chen, and G. C. Guo, Phys. Rev. Lett. {\bf 104}, 100502 (2010).
\bibitem{NM7}  A. Z. Chaudhry and J. Gong, Phys. Rev. A {\bf 85}, 012315(2012).
\bibitem{Ma-Yu2014} T. Ma, Y. Chen, T. Chen, S. Hedemann and T. Yu, Phys. Rev. A \textbf{90}, 042108 (2014).
\bibitem{Chen-Yu2014} Y. Chen, J. Q. You and T. Yu, Phys. Rev. A \textbf{90}, 052104 (2014).



\bibitem{QSD1}L.Di\'osi and W. T. Strunz, Phys. Lett. A \textbf{235}, 569 (1997).
\bibitem{QSD2}W. T. Strunz, L.Di\'osi and N. Gisin, Phys. Rev. Lett. \textbf{82}, 1801 (1999).

\bibitem{QSD3}J. Jing and T. Yu, Phys. Rev. Lett. \textbf{105}, 240403 (2010).
\bibitem{QSD4}J. Jing, X. Zhao, J. Q. You and T. Yu, Phys. Rev. A \textbf{85}, 042106 (2012).
\bibitem{QSD5}J. Jing, X. Zhao, J. Q. You, W. T. Strunz and T. Yu, Phys. Rev. A \textbf{88}, 052122 (2013).

\bibitem{NMEQ1}T. Yu, L. Di\'osi, N. Gisin, and W. T.  Strunz, Phys. Rev. A \textbf{60}, 91 (1999).
\bibitem{NMEQ2}T. Yu,   Phys. Rev. A \textbf{69}, 062107 (2004).
\bibitem{NMEQ3}S. Maniscalco and F.  Petruccione, Phys. Rev. A \textbf{73}, 012111 (2006).
\bibitem{NMEQ4}K. L. Liu and H. S.  Goan,  Phys. Rev. A \textbf{76}, 022312 (2007).


\end{thebibliography}
\end{document}